\documentclass[fleqn,10pt]{wlscirep}
\usepackage[utf8]{inputenc}
\usepackage[T1]{fontenc}
\usepackage[algoruled,algo2e]{algorithm2e}
\usepackage{sidecap}
\usepackage{boldline}
\usepackage{setspace}
\usepackage{xcolor}
\usepackage{graphicx, caption, subcaption}
\usepackage{multirow}
\usepackage{subcaption}
\usepackage{tabularx}
\usepackage{graphicx,caption}
\usepackage{amsfonts,amsmath,amssymb}
\usepackage{comment}
\usepackage{lineno}
\usepackage{hyperref}
\usepackage[explicit]{titlesec}
\usepackage{fixmath}
\usepackage[font=small,labelfont=bf,justification=justified,format=plain]{caption}

\usepackage{algorithm,algorithmic}

\newcommand{\be}{\begin{equation}}
\newcommand{\ee}{\end{equation}} 
\newcommand{\bea}{\begin{eqnarray}}
\newcommand{\eea}{\end{eqnarray}}

\renewcommand{\v}[1]{\mathbf{#1}} % for vectors
\renewcommand{\bf}[1]{\textbf{#1}} % for bold words

\newcommand{\f}[2]{\frac{#1}{#2}}
\newcommand{\ccup}[1]{\left\{#1\right\}}

\newcommand{\rup}[1]{\left[#1\right]}

\renewcommand{\t}[1]{\tilde{#1}}

\renewcommand{\ref}[1]{[\ref{#1}]}

\usepackage[algoruled,algo2e]{algorithm2e}
\usepackage{sidecap}
\usepackage{boldline}
\usepackage{setspace}
\usepackage{xcolor}

\usepackage{graphicx, caption, subcaption}
\graphicspath{{./figures/}}

\usepackage{cleveref}
\crefname{equation}{Eq.}{Eqs.}
\crefname{section}{Sec.}{Secs.}
\crefname{figure}{Fig.}{Figs.}
\usepackage[normalem]{ulem} % to use nix

\newcommand{\nrout}{\mbox{{\small ORC-Nextrout}}}
\newcommand{\otd}{\mbox{{\small OTDSinkhorn}}}
\newcommand{\mt}{\mbox{{\small MULTITENSOR}}}

\newcommand{\orcnrout}{\mbox{{\small ORC-Nextrout}}}
\title{Community Detection in networks by Dynamical Optimal Transport Formulation}

\author[1,*,+]{Daniela Leite}
\author[1,*]{Diego Baptista}
\author[1]{Abdullahi Ibrahim}
\author[2]{Enrico Facca}
\author[1]{Caterina De Bacco}
\affil[1]{Max Planck Institute for Intelligent Systems, Cyber Valley, 72076 T{\"u}bingen, Germany}
\affil[2]{ Univ. Lille, Inria, CNRS, UMR 8524 - Laboratoire Paul Painlev\'e,
  F-59000 Lille, France}

\affil[+]{daniela.leite@tuebingen.mpg.de}

\affil[*]{These authors contributed equally to this work}

\begin{abstract}
Detecting communities in networks is important in various domains of applications. While a variety of methods exists to perform this task, recent efforts propose Optimal Transport (OT) principles combined with the geometric notion of Ollivier-Ricci curvature to classify nodes into groups by rigorously comparing the information encoded into nodes' neighborhoods. We present an OT-based approach that exploits recent advances in OT theory to allow tuning for traffic penalization, which enforces different transportation schemes. As a result, our model can flexibly capture different scenarios and thus increase performance accuracy in recovering communities, compared to standard OT-based formulations. We test the performance of our algorithm in both synthetic and real networks, achieving a comparable or better performance than other OT-based methods in the former case, while finding communities more aligned with node metadata in real data. This pushes further our understanding of geometric approaches in their ability to capture patterns in complex networks.

\end{abstract}

\begin{document}

\flushbottom
\maketitle

\thispagestyle{empty}

\section{Introduction}
% !TEX root = main.tex

Complex networks are ubiquitous, hence modeling interactions between pairs of individuals is a relevant problem in many disciplines \cite{huang2021survey,newman2018networks}. Among the variety of analysis that can be performed on them, community detection \cite{fortunato2010community,fortunato2016community,weber2018detecting, samal2018comparative} is a popular application that involves finding groups (or communities) of nodes that share similar properties. The detected communities may reveal important functional properties of the underlying system. Community detection has been used in diverse areas including, discovering potential friends on social networks \cite{zhu2017emotional}, evaluating social networks \cite{wang2017sentiment}, personalized recommendation of item to user \cite{li2020personalized}, detecting potential terrorist activities on social platforms \cite{waskiewicz2012friend}, fraud detection in finance \cite{pinheiro2012community}, study epidemic spreading process \cite{chen2012epidemic} and so on. 
 
Several algorithms have been proposed to tackle this problem which utilize different approaches, such as statistical inference \cite{debacco2017community,raghavan2007near}, graph modularity \cite{clauset2004finding}, statistical physics \cite{reichardt2006statistical} and information theory \cite{rosvall2008maps}. 
Here, instead, we adopt a recent approach connecting community detection with geometry, where communities are detected using geometric methods like the Ollivier-Ricci curvature (ORC) and we exploit optimal transport theory to calculate this efficiently. 

In Riemannian geometry, a curvature quantifies how geodesic paths converge or diverge, depending on the curvature's sign. In networks, the ORC plays a similar role where edges with negative curvature are traffic bottlenecks, in terms of network flow of the shortest paths. In the opposite case, positively curved edges contribute to transport on the network along with several others, reflecting the fact that they are well connected. Defining communities as robust transport of information along with the network, we could cluster edges based on their curvature: those with positive curvature can be clustered together, while those with negative curvature may be seen as ``bridges'' connecting different communities. The idea of using Ricci curvature to find communities on networks has been recently proposed in \cite{ni2019community,sia2019ollivier}. In this work we follow a similar approach, but generalize it for the case of branched \cite{santambrogio2007optimal,facca2021branch} and congested \cite{brasco2010congested} optimal transport problems, building from recent results \cite{baptista2020network}.
Specifically, our algorithm allows us to efficiently tune the sensitivity to detecting communities in a network, by means of a parameter that controls the flow of information shared between nodes. 
We perform a comprehensive comparison between the proposed algorithm and existing ones on synthetic and real data. Our algorithm, named \orcnrout, detects communities in synthetic networks with similar or higher accuracy compared to other OT-based methods in the regime where inference is not trivial. This is also observed in a variety of real networks, where the ability to tune between different transportation regimes allows finding at least one result that outperforms other methods, including approaches based on statistical inference and modularity-based community detection.

\subsection*{Related work.} 
The idea of exploring geometrical properties of a graph, and in particular curvature, has been explored in different branches of network science, ranging from biological\cite{sandhu2015graph} to communication\cite{wang2016interference} networks.
Intuitively, the Ricci curvature can be seen as the amount of volume through which a geodesic ball in a curved Riemannian manifold deviates to the standard ball in Euclidean spaces\cite{ni2015ricci}. When defined in graphs, it indicates whether edges (those with positive values for the curvature) connect nodes inside a cluster, or if they rather bond different clusters together (those with negative values for the curvature).
	
Two main discrete graph curvature approaches have been proposed: the Ollivier-Ricci (OR) curvature based on the optimal transport theory introduced by Ollivier, \cite{ollivier2009ricci,ollivier2010survey} and Forman-Ricci curvature introduced by Forman\cite{forman2003bochner}. While the graph Laplacian-based Forman curvature is computationally fast and less geometrical, we focus on OT-based approach
due to its more geometric nature. Some applications of the Ollivier-Ricci curvature include network alignment \cite{ni2018network} and community detection \cite{sia2019ollivier, ni2019community,ye2019curvature}.
 
On the other hand, community detection in networks is a fundamental area of network science, with a wide range of approaches proposed for this task \cite{fortunato2010community,fortunato2016community,lancichinetti2009community}. 
Our work is inspired by recent OT-based methods \cite{sia2019ollivier,ni2019community} for community detection. These methods consider the OR curvature to sequentially identify and prune negatively curved edges from a network to identify communities. While our approach also considers OR curvature to prune edges, it controls the flow of information exchanged between nodes by means of parameter, making the edge pruning dynamic. This is detailed in \Cref{section2}.

\section{$\beta$-Wasserstein Community Detection Algorithm}\label{section2}

% !TEX root = main.tex

In this section,  we describe how our approach solves the community detection problem. As previously stated, we rely on optimal transport principles to find the communities. To solve the optimal transport problem applied in our analysis we use the discrete \textit{Dynamic Monge-Kantorovich} model (\textit{DMK}), as proposed by Facca et al. \cite{facca2018towards, facca2020numerical} to solve transportation problems on networks.

%% General notation 

We denote a weighted undirected graph as $G = (V,E, W)$, where $V,E,W$ are the set of nodes, edges and weights, respectively. We use the information of a node neighborhood $\mathcal{N}(i) = \ccup{j \in V | (i,j) \in E}$ to decide whether node $i$ belongs to a given community. We do this by comparing a distribution defined on $\mathcal{N}(i)$ to the ones defined on other nodes close to $i$. This distribution is defined as $m_i^{\alpha}$, where $m^{\alpha}_i(k):=	\alpha$ if $k=i$ and  $m^{\alpha}_i(k):= (1-\alpha)/|\mathcal{N}(i)| $ if $k \in \mathcal{N}(i)$. Intuitively, the distribution $m$ assigns a unit of mass to $i$ and its connections: $\alpha$ controls how much weight the node $i$ should have, and once this is assigned, its neighbors receive the remaining mass in an even way. 

\begin{figure}[h]
	\centering
	\begin{subfigure}[b]{\textwidth}
		\includegraphics[width=\textwidth]{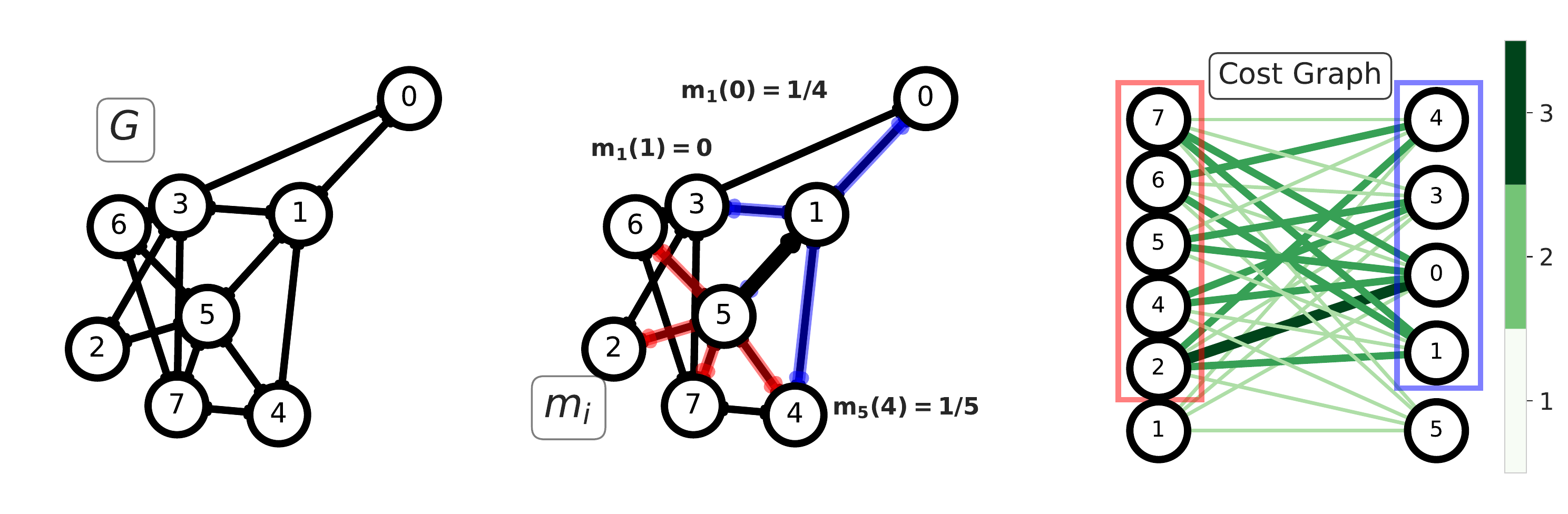}
	\end{subfigure}
	\caption{Left): an example graph $G$ where edges have unitary weights.  Center): the edge $(1,5)$ (bold black line) is selected to define the OT problem between $m_{1},m_{5}$; neighborhoods of nodes 1 and 5 are highlighted with blue and red edges and are used to build the corresponding distributions $m_{1},m_{5}$. Right): The complete bipartite graph $B_{15}$ where the OT problem is defined. The color intensity of the edges represent the distance between the associated nodes on the graph $G$, as shown by the colorbar. $m_1$ and $m_5$ are both defined for $\alpha=0,$ i.e. no mass is left in $1$ and $5$. }
	\label{fig:illustration}
\end{figure}

The next step is to compare the distribution $m_i^{\alpha}$ of the node $i$ to that of its neighbors. Consider an edge  $(i,j)\in E$ and $m_j$,  the distribution defined on the node $j$, neighbor of $i$. We assume that if $i$ and $j$ belong to the same community, then both nodes may have several neighbors in common, and therefore, $m_i$ and $m_j$ should be similar. Notice that this implies an assortativity assumption, where nodes within the same community are more likely to interact than nodes in different communities \cite{newman2018networks,fortunato2016community}. This has been observed for instance in social or biological networks \cite{asikainen2020cumulative,debacco2017community}. On the contrary, it may not be appropriate to model disassortative datasets, where nodes tend to connect more often across communities. \\
To estimate the similarity between $m_i$ and $m_j$ we use OT principles. Specifically, we compute the cost of transforming one distribution into the other. This is related to the cost of moving the mass from one neighborhood to the other, and it is assumed to be the weighted shortest-path distance between nodes belonging to $\mathcal{N}(i)$ and $\mathcal{N}(j)$. A schematic representation of the algorithm can be seen in \Cref{fig:illustration}. The OT problem is solved in an auxiliary graph, the complete bipartite network $B_{ij} = (V_{ij}, E_{ij}, \omega_{ij})$ where $V_{ij} := (V_{i},V_{j}) :=(\mathcal{N}(i) \cup \ccup{i}, \mathcal{N}(j) \cup \ccup{j})$, $E_{ij}$ is made of all the possible edges between $V_{i}$ and $V_{j}$. The weights of the edges are given by the weighted shortest path distance $d$ between two nodes measured on the input network $G$. 

%% Ollivier Ricci curvature and beta (here a reference to the impact of beta)

\begin{figure}[h]
	\centering
	\begin{subfigure}[b]{1.0\textwidth}
		\includegraphics[width=\textwidth]{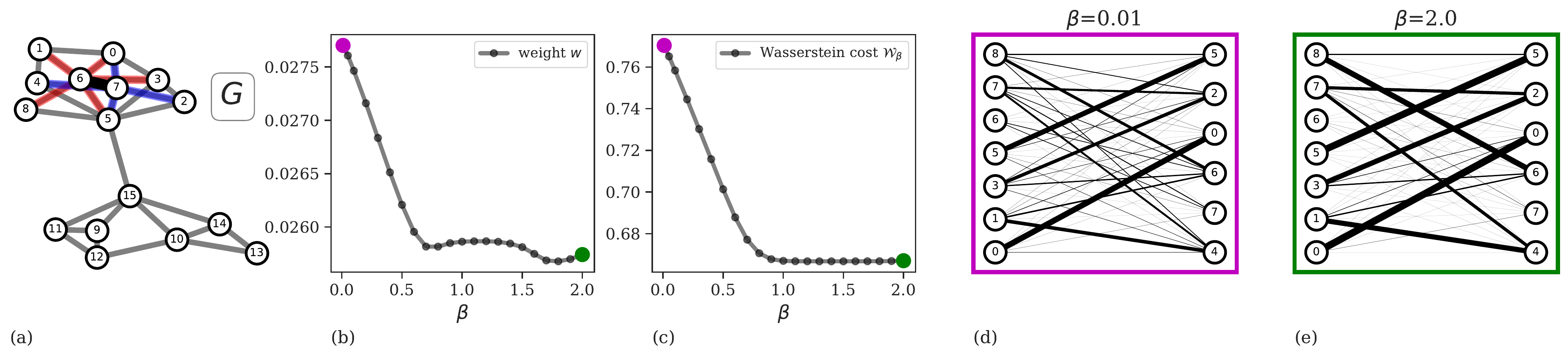}
		\label{fig:impact-of-beta-intra}
	\end{subfigure}
	\caption{Visualization of how $\beta$ impacts an intra-community edge. (a) Example intra-community structure between nodes 6 and 7. (b) The weight of edge $(6,7)$ decreases when $0<\beta<0.6$, while for $0.5<\beta<2.0$ it reaches a minimum and then slightly increases again. This justifies the better performance in detecting communities obtained for higher values of $\beta$, as shown in \Cref{fig:performance_lfr,fig:performance_sbm}. (c) A similar decreasing behavior is observed for the $\beta$-Wasserstein cost: for intra-community edges, $\beta>1$ consolidates traffic in the network as the Wasserstein cost stabilizes. (d-e) Example cost graph $B_{67}$ with fluxes solution of the OT problem (edge thickness is proportional to the amount of flux) in the regimes of small (d) and high (e) values of $\beta$.}
	\label{fig:impact-of-beta}
\end{figure}

The similarity between $m_i $ and $m_j$ is the Wasserstein cost  $\mathcal{W}(m_i,m_j,\omega_{ij})$ of the solution of the transportation problem. In its standard version, this number is the inner product between the solution $Q$, a vector of flows defined on edges, and the cost $\omega_{ij}$. In our case, since the DMK model allows to control the flow of information through a hyperparameter $\beta \in (0,2]$, we define the $\beta$\textit{-Wasserstein cost}, $\mathcal{W}_\beta(m_i,m_j,\omega_{ij})$, as the inner product of the solution $Q =Q(\beta)$ of the DMK model and the cost $\omega_{ij}$.
For $\beta=1$ we compute the Wasserstein-1 distance between $m_i$ and $m_j$, while for $\beta\ne1$ the influence of $\beta$ in the solution of the transportation problem can be seen in  \Cref{fig:impact-of-beta}. When $\beta<1,$ more edges of $B$ tend to be used to transport the mass, thus we observe congested transportation \cite{brasco2010congested}. When $\beta>1$ fewer edges are used, hence we observe branched transportation, and the $\beta$\textit{-Wasserstein cost} coincides with a branched transport distance \cite{Xia:2003,facca2021branch}. The idea of tuning $\beta$ to interpolate between various transportation regimes has been used in several works and engineering applications \cite{baptista2020network,baptista2021principled,baptista2021convergence,lonardi2021designing,lonardi2021multicommodity,lonardi2021infrastructure,ibrahim2021optimal}. 

Calculating the Wasserstein cost is necessary to determine our main quantity of interest, the discrete Olliver-Ricci curvature, defined as: 
\be
\kappa_\beta(i,j) := 1 - \dfrac{\mathcal{W}_\beta(m_i,m_j,\omega_{ij})}{d_{ij		}} \quad,
\ee
where $d_{ij}$ is the weighted shortest path distance between $i$ and $j$ as measured in $G$.
Intuitively,  if $i$ and $j$ are in the same communities, several $k \in V_{i}$ and $\ell \in V_{j}$ will be also directly connected. Thus, the Wasserstein distance between $m_{i}$ and $m_{j}$ will be shorter than $d_{ij}$, yielding a positive $\kappa_\beta(i,j)$. Instead, when $i$ and $j$ are in different communities, their respective neighbors will be unlikely connected, hence $d(i,j)<\mathcal{W}_\beta(m_i,m_j,\omega_{ij})$, yielding a negative $\kappa_\beta(i,j)$.  

%% Ricci flow update and beta

The Ricci flow algorithm on a network is defined by iteratively updating the weights of the graph $G$ \cite{ni2019community,sia2019ollivier}. These are updated by combining the curvature and shortest path distance information \cite{ollivier2009ricci}. We redefine these updates using our proposal for the Ollivier-Ricci curvature:
\be
w_{ij}^{(t+1)} := d_{ij}^{(t)} - \kappa_\beta^{(t)}(i,j)\cdot d_{ij}^{(t)},
\ee
where $w_{ij}^{(t+1)}$ is the weight of edge $(i,j)$ at time $t$,  $w_{ij}^{(0)} = d_{ij}^{(0)}, $ and $d_{ij}^{(t)}$ is the shortest path distance between nodes $i$ and $j$ at iteration $t$. At every time step $t$, the weights are normalized by their total sum.

The algorithm \orcnrout {} dynamically changes the weights of the graph $G$ to isolate communities: intra-community edges will be shortened, while inter-community ones will be enlarged. These changes are reached after different number of iterations,  depending on the input data. We choose the one that maximises some predefined quality measure. To find the communities we apply a \textit{network surgery} criterion as proposed by Ni et al. \cite{ni2019community} based on the stabilisation of the modularity of the network. Notice that our algorithm  does not need prior information about the number of communities: edges will be either enlarged or shortened depending on the optimal transport principles agnostic to community labelings. 

 A pseudo-code of the implementation is shown in Algorithm \ref{alg:w}. 
\begin{algorithm}[htb]
   \caption{\orcnrout {}}
   \label{alg:w}
\begin{algorithmic}
  \STATE {\bfseries Input:} $G=(V,E,W),$ traffic rate $\beta,$ \textit{MaxIterNum}$\in \mathbb{N}$\\
    \STATE {\bfseries Output:} updated $W$ \\  
    \STATE { Initialize: $\mathbf{w}^0=W$}
    \FOR { $t \in range(MaxIterNum)$} 
         \FOR {$e=(i,j)\in E\;\; $}
         \STATE{ Calculate $m_i, m_j$ }\\
         \STATE{ Build $B_{ij}$}\\
         \STATE{ Get $Q(\beta) \in \mathbb{R}^{|E_{ij}|},$  $Q(\beta)= DMK(B_{ij},m_i,m_j,\beta)$}
         \STATE{ Compute $\kappa_\beta(e)$}
         \STATE{ Compute $\mathbf{w}_\beta(e)$}
		 \ENDFOR
	\STATE{ Update $\mathbf{w}^t = \mathbf{w}_\beta$}
	\ENDFOR
\end{algorithmic}
\end{algorithm}

\section{Results on Community Detection problems}\label{section4}

\subsection*{Synthetic Networks}
To investigate the accuracy of our model in detecting communities, we consider synthetic networks generated using the \textit{Lancichinetti–Fortunato–Radicchi} (LFR) benchmark \cite{lancichinetti2008benchmark} and the \textit{Stochastic Block Model} (SBM) \cite{holland1983stochastic}. Both models provide community labels used as \textit{ground-truth} information during the classification tasks. 

\begin{figure}[h!]
	\centering

	\begin{subfigure}[b]{1.0\textwidth}
	%\caption{LFR}
	\includegraphics[trim =65mm 10mm 65mm 3mm, clip, width=1.0\textwidth]{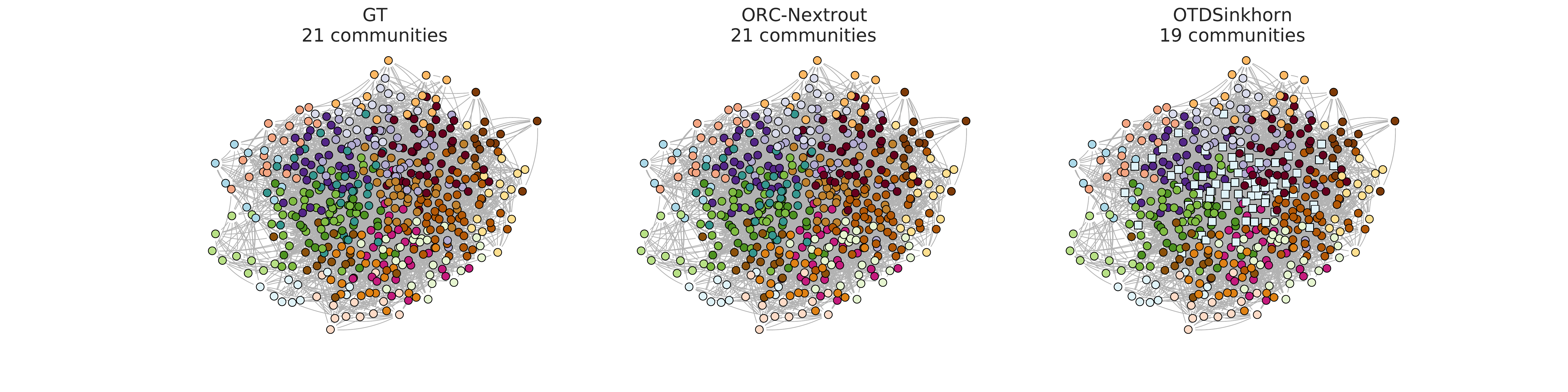}
	\label{fig:comm-structure-lfr}
	\end{subfigure}
	\caption{Example of community structure on a synthetic LFR network. The rightmost panel shows the ground-truth community structures to be predicted in an LFR network generated using $\mu = 0.35$. Square-shaped markers denote nodes that are assigned to communities different than those in ground-truth.  In middle and last panels, \nrout{}  with $\beta=2$ perfectly retrieves the 21 communities, while \otd{} predicts only $19$ communities with an ARI score of 0.73, wrongly assigning ground-truth dark green and light brown (square-shaped) nodes to the light blue community. }
	\label{fig:sbm-example-gt}
\end{figure}

%LFR
\textit{Lancichinetti–Fortunato–Radicchi benchmark:} 
this benchmark generates undirected unweighted networks $G$ with disjoint communities. It samples node degrees and community sizes from power law distributions, see \Cref{fig:sbm-example-gt} for an example. One of its advantages is that it generates networks with heterogeneous distributions of degrees and community sizes. The main parameters in input are the number of nodes $N,$ two exponents $\tau_1$ and $\tau_2$ for the power law distributions of the node degree and community size respectively, the expected degree $d$ of the nodes, the maximum number of communities on the network $K_{max}$ and a fraction $\mu$ of inter-community edges incident to each node. To test the performance of our algorithm, we use the set of LFR networks used and provided by the authors of \cite{ni2019community}.  We set $\tau_1=2,$ $\tau_2 = 1,$ $d=20,$  $K_{max}=50$ and $\mu \in [0.05,0.75]$.

%SBM
\textit{Stochastic Block Model:}
this model probabilistically generates networks with non-overlapping communities. One specifies the number of nodes $N$ and the number of communities $K$, together with the expected degree $d$ of a node and a ratio $r\in[0,1]$. Networks are generated by connecting nodes with a probability $r*p_{intra}$ if they belong to different communities; $p_{intra}$ if they are part of the same community, where $p_{intra} = d\times K / N$. Notice that the smaller the ratio $r$ is, the less inter-community connections would exist, which leads to networks with a more distinct community structure. 

We set $N=500$, $K=3$, $d=15$ and $r\in[0.01, 0.5]$ and generate 10 random networks per value of $r$. 

\paragraph{Results.}
To evaluate the performance of our method in recovering the communities, we use the \textit{Adjusted Rand Index} (ARI) \cite{hubert1985comparing}. ARI compares the obtained community partition with the \textit{ground truth} clustering. It takes values ranging from 0 to 1,  where ARI=0 is equivalent to random community assignment, and ARI=1 denotes perfect matching with the ground truth communities, hence the higher this value the better the recovery of communities.

We test our algorithm for different types of information spreading in our OT-based model, as controlled by the parameter $\beta$, using the software developed in \cite{facca2021iterative} \footnote{Source code at  \href{https://gitlab.com/enrico_facca/dmk_solver}{https://gitlab.com/enrico\_facca/dmk\_solver}}. We used $\beta = 1$, i.e., standard Wasserstein distance; $\beta \in \{0.1,0.5\}$ for congested transportation, enforcing broad spreading across the neighbors; and $\beta \in \{1.5,2 \}$ to favor branching schemes, where fewer edges are used to decide which community a node should belong to. For the OT-based algorithms, we run 15 iterations and choose the one with the best ARI scores. In some cases, high scores are reached in fewer iterations. 

\begin{figure}[h]
	\centering	
	
	\begin{subfigure}[b]{{.45\linewidth}}
	\captionsetup{justification=centering}
	\includegraphics[width=\textwidth]{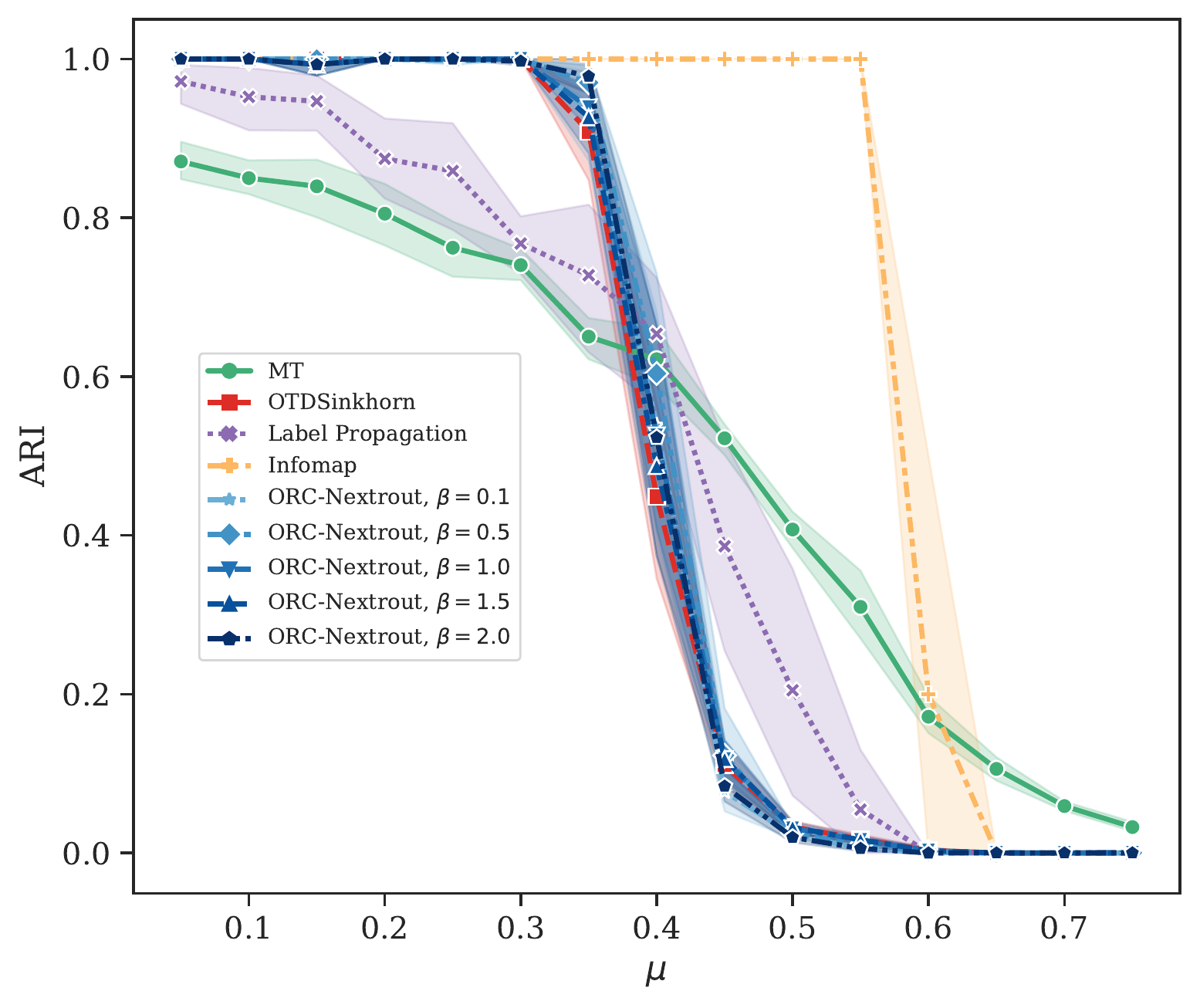}
	\caption{LFR} 
	\label{fig:performance_lfr}
	\end{subfigure}
	\begin{subfigure}[b]{{.45\linewidth}}
	\captionsetup{justification=centering}
	\includegraphics[width=\textwidth]{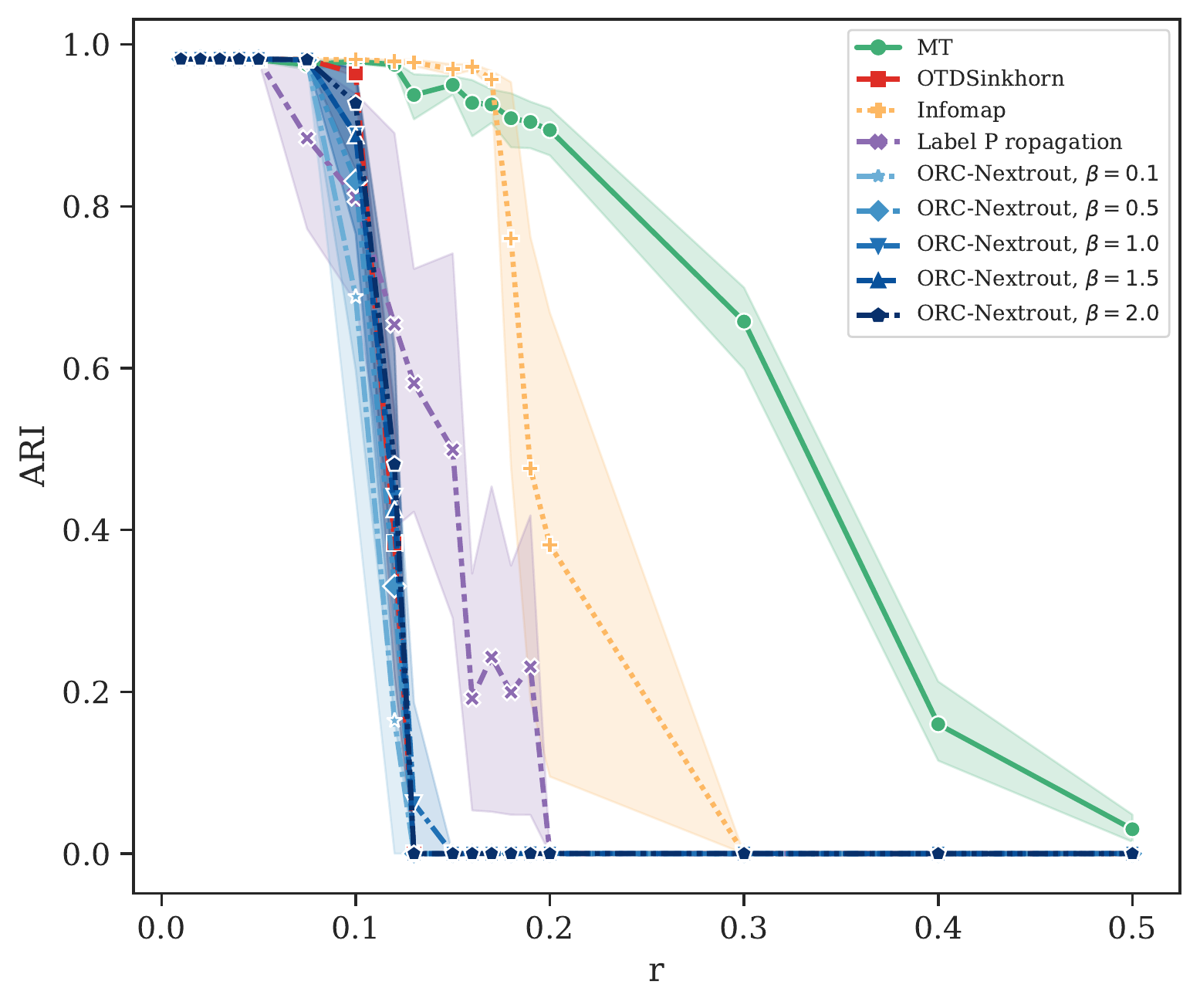}
	\caption{SBM}\label{fig:performance_sbm}
	\end{subfigure}
	\caption{Results on LFR and SBM synthetic data. Performance in detecting ground-truth communities is measured by the ARI score. Markers and shadows are the averages and standard deviations  over 10 network realisations with the same value of the parameter used in generation. Markers' shape denote different algorithms. a) LFR graph with $N=500$ nodes and different values of $K$ ranging from $(17,22)$.
	b) SBM with $N=500$ nodes, $K=3$ communities and average degree $d=15$. The parameter $r$ is the ratio of inter-community with intra-community edges. }
	\label{fig:results-syntetic}
\end{figure}

The results in \Cref{fig:results-syntetic} show the performance in both LFR and SBM benchmarks with OT-based methods, our method for various $\beta$ and one based on the Sinkhorn algorithms (\otd) \cite{cuturi2013sinkhorn,flamary2021pot }. Our main goal is to assess the impact of tuning between different transportation regimes (as done by $\beta$) in terms of community detection via OT principles. Nevertheless, to better contextualize the performance of OT-based algorithms in the wide spectrum of community detection methods,  we also include comparisons with algorithms that are not OT-based. Namely,  we consider  a probabilistic model with latent variables (MT)\cite{debacco2017community}, and with two modularity-based algorithms, Label Propagation\cite{raghavan2007near} and Infomap\cite{rosvall2008maps}. Our algorithm outperforms \otd \text{} for various values of $\beta$ in an intermediate regime where OT-based inference is not trivial. This happens in both benchmark LFR and SBM, as shown in \Cref{fig:results-syntetic}. For lower and higher values of the parameters, performance is similar and close to the two extremes of ARI = 0 and 1. OT-based methods have a similar sharp decay in performance from the regime where inference is easy to the more difficult one, as also observed in \cite{ni2019community}. The other community detection methods have smoother decay, but with lower performance in the regime where OT-based approaches strive, except for Label Propagation and MT, which are more robust in this sense. In the intermediate regime where inference is not trivial (i.e. along the sharp decay of OT-based methods), we observe that different values of $\beta$ give higher performance than \otd. For SBM the highest performance is achieved consistently for high $\beta=2$, while for LFR the best $\beta$ varies with $\mu$. A qualitative  example where  \nrout{} \text{} is performing better than \otd \text{},  in an instance of LFR of this intermediate regime, is shown in \Cref{fig:sbm-example-gt}.

\subsection*{Analysis of real networks}

Next, we evaluate our model on various real datasets\cite{networkdata} containing node metadata that can be used to assess the recovery of communities. While failing to recover communities that align well with node metadata should not by automatically interpreted as a model's failure \cite{peel2017ground} (e.g. the inferred communities and the chosen node metadata may capture different aspect of the data), having a reference community structure to compare against allows to inspect quantitatively difference between models. These real networks differ on structural features like number of nodes, average degree, number of communities and other standard network properties as detailed in Table \ref{tab:data_desc}.  
Specifically, we consider i) a network of co-appearances of characters in the novel \textit{Les Misérables} \cite{knuth1993stanford} (Les Miserables). Edges are built between characters that encounter each other. ii) A network of $62$ bottlenose dolphins in a community living off Doubtful Sound, in New Zealand \cite{lusseau2003bottlenose} (Dolphins). Nodes represent dolphins, and edges indicate frequent associations between them. This network is clustered into four groups, conjectured as clustered from one population and three sub-populations based on the interactions between dolphins of different sex and ages \cite{lusseau2004identifying}. The dolphins were observed between 1994 and 2001. iii) A network of Division I matches of American Football during a regular season in the fall of 2000 \cite{girvan2002community} (American football). Nodes represent teams and edges are games between teams. Teams can be clustered according to their football college conference memberships. iv) A network of books on US politics published around the $2004$ presidential election and sold by an online bookseller\cite{politicalbooks} (Political books). Nodes represent the books and the edges between books are frequent co-purchasing of books by the same buyers. Books are clustered based on their political spectrum as neural, liberal or conservative. 
 
\begin{table}[htbp]
	\begin{center}
		\caption{\bf {Real networks description.} We report statistics for the real networks used in our experiments. \textit{N} and \textit{E} denote the number of nodes and edges, respectively. $K$ is the number of communities in the ground truth data. AvgDeg, AvgBtw and AvgClust are the average degree, betweenness centrality and average clustering coefficient, respectively.}
		%\begin{adjustbox}{angle=0}
		\resizebox{0.7\columnwidth}{!}{%
			{\renewcommand{\arraystretch}{1.11}
				\begin{tabular}{lllllll}
					\toprule
					\textbf{Dataset} & $N$& $E$ &$K$  &\textbf{AvgDeg}  & \textbf{AvgBtw} & \textbf{AvgClust}  \\
					\midrule
					Les Miserables & $77$& $254$ & $11$ & $6.6$ & $0.0219$ & $0.5731$  \\
					Dolphins & $62$& $159$ & $4$ &$5.1$&$0.0393$&   $0.2590$ \\
					American football &  $115$&$613$&$12$ &$10.7$ &$0.0133$&   $0.4032$  \\
					Political books &  $105$&$441$&$3$ &$8.4$ &$0.0202$&  $0.4875$  \\
					\bottomrule
				\end{tabular}%
		}} 
		%\end{adjustbox}
		\label{tab:data_desc}
	\end{center}
\end{table}

\begin{figure}[h]
	\centering
		\includegraphics[width=\textwidth]{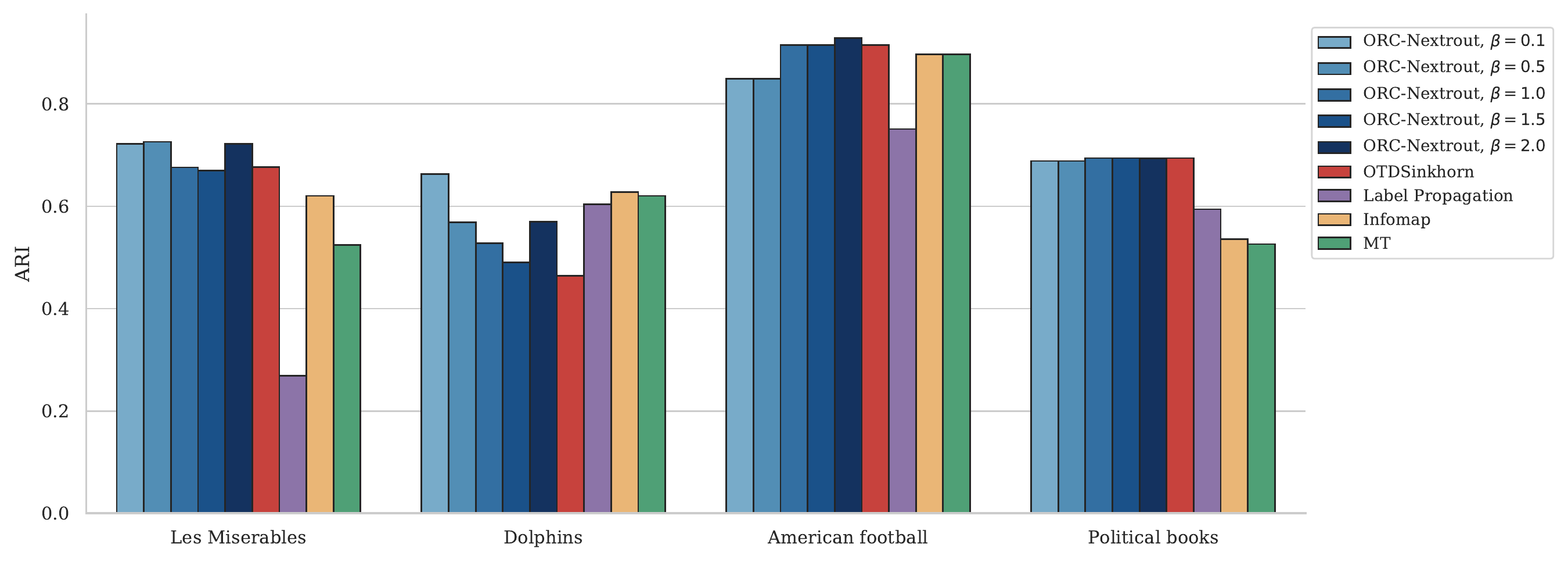}
	\caption{Results on real data. Performance in terms of recovering communities using metadata information is calculated in terms of the ARI score. \orcnrout { } shows competing results  against all methods with different optimal $\beta$ across datasets.}
	\label{fig:res-on-real}
\end{figure}

\begin{figure}[htbp]
	\begin{subfigure}[b]{\textwidth}
		\centering
		\caption{Les Miserables}
		\includegraphics[trim = 100mm 15mm 80mm 1mm, clip, width=1.0\textwidth]{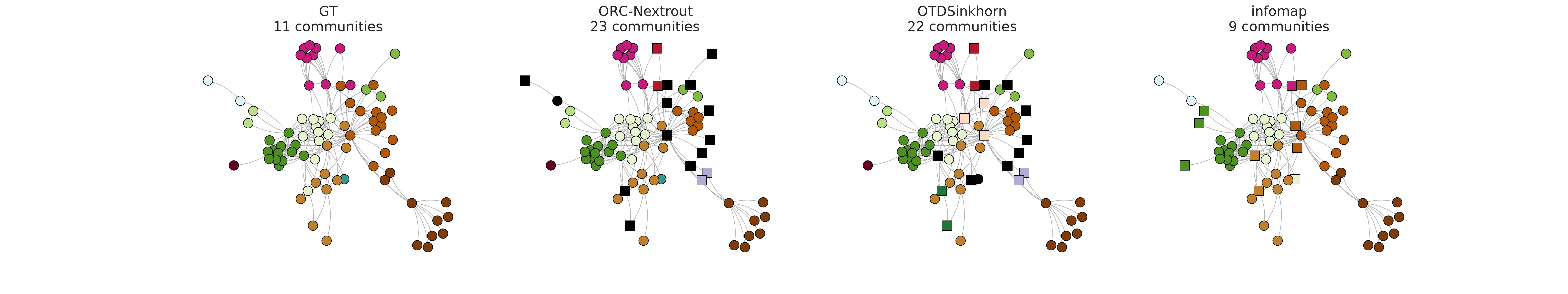}
		\label{fig:comm-les-mis}
	\end{subfigure}

	\begin{subfigure}[b]{\textwidth}
	\centering
	\caption{Political books}
	\includegraphics[trim = 100mm 15mm 80mm 1mm, clip, width=1.0\textwidth]{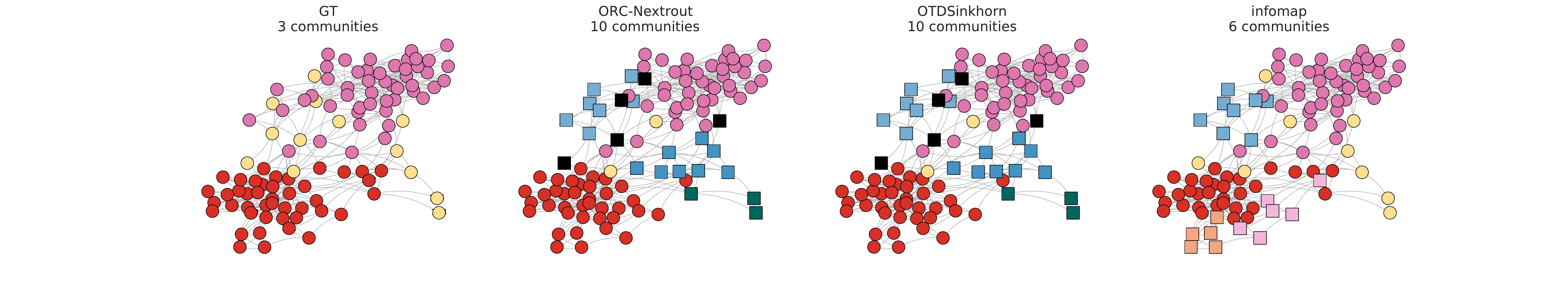}
	\label{fig:comm-political-books}
	\end{subfigure}

	\caption{Communities in real networks. We show the communities inferred by \nrout {} ($\beta=0.5,1.5$ for top and bottom rows respectively), \otd \text{} and Infomap and compare against those extracted using node attributes (GT). The visualization layout is given by the \textit{Fruchterman-Reingold force-directed}  algorithm \cite{fruchterman1991graph}, therefore, groups of well-connected nodes are located close to each other. Dark nodes represent individual nodes who are assigned to isolated communities by OT-based methods. Square-shaped markers denote nodes assigned to communities different than those obtained from node metadata. }
	\label{fig:realnets-plots}
\end{figure}

OT-based algorithms outperform other community detection algorithms in detecting communities aligned with the node metadata, as shown in \Cref{fig:res-on-real}. In particular, \nrout \text{} has the highest accuracy performance considering the best performing  $\beta$. The impact of tuning this parameter is  noticeable from these plots, as the best performing value varies across datasets. In the Les Miserables and Dolphins networks, $\beta<1$ has better performance, while in American Football the best performing value is for $\beta>1$. Performance is similar across OT-based methods in the Political Books network. 
In \Cref{fig:realnets-plots} we show the communities detected by the best performing \orcnrout {} version together with \otd \text{} and Infomap in Les Miserables and Political books. Focusing on Les Miserables, we see how  \orcnrout {} successfully detects three characters in the green communities, in particular a highly connected node in the center of the figure (in dark green). Notice that these are placed in the same (pink or black) community by \otd.  Thus  \orcnrout {} achieves a higher ARI than \otd.  Both OT-based approaches retrieve well communities exhibiting clustering patterns, with many connections within community. Instead, they both divide the communities with a hub and spokes structure due to the lack of common connections within the group.

The communities detected in the Political books datasets highlight the tendency of OT-based methods to extract a larger number of communities (10) than those observed from node metadata (3). Among these extra communities, 3 are made of a few nodes, while 5 of them are made of one isolated node each. This is related to the fact that OT-based methods perform particularly well for networks with internally densely-connected community structures, but may be weaker for community structures that are sparsely connected \cite{sia2019ollivier}. One could potentially assign these nodes to larger communities, for instance by preferential attachment as done in \cite{sia2019ollivier}, thus in practice reducing the number of communities. Devising a principled method or criterion to do this automatically is an interesting topic for future work. 
This tendency is further corroborated by the fact that OT-based algorithms recover robustly the two communities that are mostly assortative (red and pink in the figure), while they struggle to recover the disassortative community depicted in the centre (yellow). This is community has several connections with nodes in the other two communities and has been separated into smaller groups by OT-based approaches, as described above. This also highlights the need for methods that are robust against situations where a combination of assortative and disassortative communities coexist in a network.   \\

\section*{Conclusion}\label{section5}
Community detection on networks is a relevant and challenging open area of research. Several methods have been proposed to tackle this issue, with no “best algorithm” that fits well every type of data. We focused here on a recent line of work that exploits principles from Optimal Transport theory combined with the geometric concept of Ollivier-Ricci curvature applied to discrete graphs. Our method is flexible in that it tunes between different transportation regimes to extract the information necessary to compute the OR curvature on edges. On synthetic data, our model is able to identify communities more robustly than other OT-based methods based on the standard Wasserstein distance in the regime where inference is not trivial. On real data, our model shows either better or comparable performance in recovering community structure aligned with node metadata compared to other approaches, thanks to the ability to tune the parameter $\beta$. \\
A relevant advantage of OT-based methods is that the number of communities is automatically learned from data, contrarily to other approaches that need this as an input parameter. In this respect, our model has the tendency of overestimating this number, similarly to other OT-based methods. Understanding how to properly incorporate small-size communities into larger ones in a principled and automatic way is an interesting topic for future work. Similarly, it would be interesting to quantify the extent to which various $\beta$ capture different network topologies. To address this, one could for instance use methods to calculate the structural distance between networks \cite{xiao2021deciphering} and correlate this against the values of the best performing $\beta$. \\
There are a number of directions in which this model could be extended.
Nodes can be connected in more than one way, as in multilayer networks. Our model could be extended by considering a different $\beta$ for each edge type, as done in \cite{ibrahim2021optimal}.
 Similarly, real networks are often rich in additional information, e.g. attributes on nodes. It would be interesting to incorporate a priori additional information to inform community detection \cite{contisciani2020community,newman2016structure}. This information can potentially be used to mitigate the problem of overestimation of the number of communities, as explained above.

\section*{Methods}\label{section_methods}

\subsection*{Optimal Transport Formulation} 

Consider the proabability distributions  $q$ that take pairs of vertices and also satisfy the constraints $\sum_i q_{ij} = m_j, \sum_j \, q_{ij} = m_i$.  In other words, these are the joint distributions whose marginals are $m_i$ and $m_j$. We call these distributions   \textit{transport plans} between $m_i$ and $m_j$. The Optimal Transport problem we are interested in is that of finding a transport plan $q^*$ that minimises the quantity $\sum_{i\sim j} q_{ij}d_{ij} ,$ where $i\sim j$ means that nodes $i$ and $j$ are neighbors and $d_{ij}$ is the cost of transporting mass from $i$ to $j$, e.g. the distance between these two nodes. The quantity $ \mathcal{W}_\beta(m_i,m_j,d):= \sum_{i,j} q^*_{ij} \,d_{ij}$, defined for this optimal $q^*$, is the \textit{Wasserstein distance} between $m_i$ and $m_j$.

\subsection*{The Dynamical Monge-Kantorovich model} 

It was recently proved \cite{facca2018towards, facca2020numerical} that solutions of the optimal transport problem previously stated can be found by turning that problem into a system of differential equations. This section is dedicated to describe this dynamical formulation.

Let $G=(V,E,W)$ be a weighted graph, with $N$ the number of nodes and $E$ the number of edges in $G$.  Let  $\mathbf{B}$ be the \textit{signed incidence} matrix of $G$. 
Let  $f^+$ and $f^-$ be two $N$-dimensional discrete distributions such that $\sum_{i\in V}f_{i}=0$ for $f = f^+-f^-$; let $\mu(t)\in \mathbb{R}^{E}$ and $u(t)\in \mathbb{R}^{N}$ be two time-dependent functions defined on edges and nodes, respectively. The discrete \textit{Dynamical Monge-Kantorovich model} can be written as:

\bea\label{eqn:discrete-DMK1}
f_{i}  &=&\sum_{e} B_{ie} \f{\mu_{e}(t)}{w_e} \sum_{j}B_{ej} \, u_{j}(t) \,, \label{eqn:kirk-discrete}\\ %\MP{\qquad\text{ or }\qquad \mathbf{B}\left(\mathbf{M}(t)\mathbf{L}^{-1}\right)\mathbf{B}^Tu(t)=f} 
\mu_{e}'(t)&=&\rup{\f{\mu_{e}(t)}{w_e}\left|\sum_{j}B_{ej}\,u_{j}(t)\right|}^{\beta}-\mu_{e}(t)  \label{eqn:mu-discrete}\,,\\ %=\rup{\mu_{e}(t)\bup{|\mathbf{L}^{-1}\mathbf{B}^Tu(t)|}_{e}}^{\beta_{d}}-\mu_{e}(t)\\
\mu_{e} (0)&>&0 \,,\label{eqn:IC-discrete}
%\mu'(t)&=& |\mathbf{C}[\mu(t)]   \nabla u(t)|-\mu(t)
\eea
where $|\cdot|$ is the absolute value element-wise. Equation (\ref{eqn:kirk-discrete}) corresponds to Kirchhoff's law, Eq. (\ref{eqn:mu-discrete}) is the discrete dynamics with $\beta$ a traffic rate controlling the different routing optimization mechanisms;  Eq. (\ref{eqn:IC-discrete}) is the initial distribution for the edge conductivities.

For $\beta = 1$ the dynamical system described by Eqs. (\ref{eqn:kirk-discrete})-(\ref{eqn:IC-discrete}) is known to reach a steady state, i.e., the updates of $\mu_e$ and $u_e$ converge to  stationary functions $\mu^*$ and $u^*$ as $t$ inscreases. The flux function $q$ defined as $q^*_{e}: = \mu^*_{e} |u^*_{i}-u^*_{j}|/w_e$ is the solution of the optimal transport problem presented in the previous section.  Notice that $\mu$ and $u$ depend on the chosen traffic rate $\beta$, and thus, so does $q=q(\beta)$. Therefore we can introduce a generalized version of the distance $\mathcal{W}$:
$$
	\mathcal{W}_\beta(m_i,m_j,w) := \sum_{i,j} q^*_{ij}(\beta)\,\,w_{ij}.
$$
We then redefine the proposed Ollivier-Ricci curvature as:
$$
\kappa_\beta(i,j) := 1 - \dfrac{\mathcal{W}_\beta(m_i,m_j,w)}{d_{ij}}.
$$

\subsection*{Probability distributions on neighborhoods}

\orcnrout {} takes in input a graph and a forcing term. While the graph encapsulates the neighborhood information provided by the nodes $i$ and $j$, the forcing function is related to the distributions one needs to transport. Analogously to what proposed by \cite{ni2019community}, we define this graph to be the weighted \textit{complete bipartite} $B_{ij} = (V_{ij}, E_{ij}, \omega_{ij})$. The weights in $\omega_{ij}$  change iteratively based on the curvature. Notice that a bipartite graph must satisfy $\mathcal{N}(i)\cap \mathcal{N}(j) = \varnothing,$ which does not hold true if $i$ and $j$ have common neighbors (this is always the case since $i\in \mathcal{N}(j)$). Nonetheless, this condition does not have great repercussions in the solution of the optimal transport problem since the weights corresponding to these edges (of the form $(i,i)$) are equal to 0. As for the forcing function, we define it to be $f := f^+ - f^- = m_i - m_j$.

\subsection*{Other methods}\label{other_methods}

% !TEX root = main.tex

To evaluate the performance of \orcnrout, we compare with some of the well-established community detection algorithms including: Infomap\cite{rosvall2008maps}, MULTITENSOR\cite{debacco2017community} (MT),  discrete Ricci flow\cite{ni2019community} (\otd), and Label propagation\cite{raghavan2007near}. We briefly describe each of these algorithms as follows;

\begin{itemize}
	\item  The \textit{Discrete Ricci flow} (here addressed as \otd) \cite{ni2019community} is an iterative node clustering algorithm that deforms edge weights as time progresses, by shrinking sparsely traveled links and stretching heavily traveled edges. These edge weights are iteratively updated based on neighborhood transportation Wasserstein costs, in a similar way to what proposed in this manuscript. After a predefined number of iterations, heavily traveled links are removed from the graph. Communities are then obtained as the connected components of this modified network.
	\item \mt \text{} (MT) \cite{debacco2017community} is an algorithm to find communities in multilayer networks. It is a probabilistic model  with latent variables regulating community structure and runs with a complexity of $O(E K)$ with assortative structure (as we consider here), where $K$ is the number of communities. This model assumes that the nodes inside the communities can belong to multiple groups (mixed-membership). In this implementation we use their validity for single layer networks (a particular case of a multilayer network). 
	\item \textit{Infomap}\cite{rosvall2008maps} employs information theoretic approach for community detection. This method uses the map equation to attend patterns of flow on a network. This flow is simulated using random walkers' traversed paths. Based on the theoretic description of these paths,  nodes with quick  information flow are then clustered into the same groups. The algorithm runs in $O(E)$.
	\item \textit{Label propagation}\cite{raghavan2007near} assigns each node to same community as majority of its neighbors. Its working principle start by initializing each node with a distinct label and converges when every node has same label as majority of its neighboring node. The algorithm has a complexity scaling as $O(E)$.
\end{itemize}

\section*{Data Availability}
The real data can be obtained from network data \cite{networkdata} and the synthetic one from the corresponding author upon request.

\bibliographystyle{apsrev4-1}
\bibliography{bibliography}
%%%%%%%%%% Acknowledgements
\section*{Acknowledgements}
The authors thank the International Max Planck Research School for Intelligent Systems (IMPRS-IS)
for supporting Daniela Leite,  Diego Baptista and Abdullahi Ibrahim.

\section*{Author contributions statement}
All authors contributed to developing the models, conceived the experiments, analyzing the results and reviewing the manuscript. DL, DB and AI conducted the experiments.
\section*{Additional information}
\textbf{Accession codes}:  open source codes and executables are available at \href{https://github.com/Danielaleite/ORC-Nextrout}{https://github.com/Danielaleite/ORC-Nextrout}. \\%(for \orcnrout) and at \href{https://gitlab.com/enrico_facca/dmk_solver}{https://gitlab.com/enrico\_facca/dmk\_solver}. \\
\textbf{Competing interests}. The authors declare no competing interests.

\newcommand{\beginsupplement}{%
        \setcounter{table}{0}
        \renewcommand{\thetable}{S\arabic{table}}%
        \setcounter{figure}{0}
        \renewcommand{\thefigure}{S\arabic{figure}}%
        \setcounter{equation}{0}
        \renewcommand{\theequation}{S\arabic{equation}}
         \setcounter{section}{0}
        \renewcommand{\thesection}{S\arabic{section}}
 }

\clearpage
\beginsupplement
%\begin{widetext}

\end{document}